%% file: BESIII_draft_BF_K0barev_2pi0.tex
\documentclass[cpc,amsmath,showpacs,twocolumn]{revtex4-1}
\usepackage{graphicx}
\usepackage{color}
\usepackage[mathlines]{lineno}

\begin{document}

\title{\boldmath Measurement of the absolute branching fraction of $D^{+}\rightarrow\bar K^0 e^{+}\nu_{e}$
via $\bar K^0\to\pi^0\pi^0$}

\input{BESIII_authors_en}

\begin{abstract}
By analyzing 2.93 fb$^{-1}$ data collected at the center-of-mass
energy $\sqrt s=3.773$ GeV with the BESIII detector, we measure the
absolute branching fraction of the semileptonic decay
$D^+\rightarrow\bar K^0 e^{+}\nu_{e}$ to be ${\mathcal
B}(D^{+}\rightarrow\bar K^0 e^{+}\nu_{e})=(8.59 \pm 0.14 \pm
0.21)\%$ using $\bar K^0\to K^0_S\to \pi^0\pi^0$, where the first
uncertainty is statistical and the second systematic. Our result is
consistent with previous measurements within uncertainties.
\end{abstract}


\maketitle

\section{Introduction}

The study of semileptonic decays of $D$ mesons can shed light on the
strong and weak effects in charmed meson decays. The absolute
branching fraction ${\mathcal B}$ of the semileptonic decay $D^+\to
\bar K^0 e^+\nu_e$ can be used to extract the form factor $f^K_+(0)$
of the hadronic weak current or the quark mixing matrix element
$|V_{cs}|$~\cite{rongg}, which are important to calibrate the
lattice quantum chromodynamics calculation on $f^K_+(0)$ and to test
the unitarity of the quark mixing matrix. In addition, the measured
${\mathcal B}(D^+\to \bar K^0 e^+\nu_e)$ can also be used to test
isospin symmetry in the $D^+\to \bar K^0e^+\nu_e$ and $D^0\to
K^-e^+\nu_e$ decays~\cite{mark3,bes2-k0ev,cleo-k0ev}. Therefore,
improving the measurement precision of ${\mathcal B}(D^+\to \bar K^0
e^+\nu_e)$ will be helpful to better understand the $D$ decay
mechanisms.

Measurements of ${\mathcal B}(D^{+}\rightarrow\bar K^0e^{+}\nu_{e})$
via $\bar K^0\to K^0_S\to\pi^+\pi^-$ have been performed by the
MARKIII, BES, CLEO and BESIII
Collaborations~\cite{mark3,bes2-k0ev,cleo-k0ev,bes3-k0ev2}.
Recently, a measurement of ${\mathcal
B}(D^{+}\rightarrow\bar K^0_Le^{+}\nu_{e})$ has been carried out by
the BESIII Collaboration~\cite{bes3-k0ev}. However, no measurement
of ${\mathcal B}(D^{+}\rightarrow\bar K^0e^{+}\nu_{e})$ using $\bar
K^0\to K^0_S\to\pi^0\pi^0$ has been reported so far. As a first
step, we present in this paper a measurement of
${\mathcal B}(D^{+}\rightarrow\bar K^0e^{+}\nu_{e})$ using $\bar
K^0\to K^0_S\to\pi^0\pi^0$, based on an analysis of 2.93~$\rm
fb^{-1}$ of $e^+e^-$ collision data~\cite{BESIII292} accumulated at
the center-of-mass energy $\sqrt s= 3.773\;\text{GeV}$ with the
BESIII detector~\cite{BESIII}. Since there is currently no room to
improve our measurement of $f^K_+(0)|V_{cs}|$~\cite{bes3-kev}, we
only aim to measure the ${\mathcal B}(D^{+}\rightarrow\bar
K^0e^{+}\nu_{e}$) in this work.

\section{BESIII detector and Monte Carlo}
The BESIII detector is a cylindrical detector with solid-angle 93\%
of $4\pi$ that operates at the BEPCII collider. It consists of
several main components. A 43-layer main drift chamber (MDC)
surrounding the beam pipe performs precise determinations of charged
particle trajectories and provides ionization energy loss ($dE/dx$)
measurements that are used for charged particle identification
(PID). An array of time-of-flight counters (TOF) is located radially
outside the MDC and provides additional charged particle
identification information. A CsI(Tl) electromagnetic calorimeter
(EMC) surrounds the TOF and is used to measure the energies of
photons and electrons. A solenoidal superconducting magnet located
outside the EMC provides a 1~T magnetic field in the central
tracking region of the detector. The iron flux return of the magnet
is instrumented with about $1272\;\text{m}^2$ of resistive plate
muon counters (MUC) arranged in nine layers in the barrel and eight
layers in the endcaps that are used to identify muons with momentum
greater than $0.5\;\text{GeV}/c$. More details about the BESIII
detector are described in Ref.~\cite{BESIII}.

A GEANT4-based~\cite{geant4} Monte Carlo (MC) simulation software,
which includes the geometric description and a simulation of the
response of the detector, is used to determine the detection
efficiency and to estimate the potential backgrounds. An inclusive
MC sample, which includes generic $\psi(3770)$ decays, initial state
radiation (ISR) production of $\psi(3686)$ and $J/\psi$, QED ($e^+
e^- \to e^+ e^-$, $\mu^+ \mu^-$, $\tau^+\tau^-$) and $q\bar
q~(q=u,d,s)$ continuum processes, is produced at $\sqrt
s=3.773\;\text{GeV}$. The MC events of $\psi(3770)$ decays are
produced by a combination of the MC generators KKMC \cite{kkmc} and
PHOTOS \cite{photons}, in which the effects of ISR \cite{isr} and
Final State Radiation (FSR) are considered. The known decay modes of
charmonium states are generated using EvtGen \cite{evtgen} with the
branching fractions taken from the Particle Data Group
(PDG)~\cite{pdg2010}, and the remaining events are generated using
LundCharm \cite{lundcharm}. The $D^+ \to \bar K^0 e^+\nu_e$ signal
is modeled by the modified pole model~\cite{BK-model}.

\section{Measurement}

\subsection{Single tag $D^{-}$ mesons}
\label{sec:evtsel}

With a mass of $3.773\;\text{GeV}$ just above the open charm
threshold, the $\psi(3770)$ resonance decays predominately into
$D^0\bar{D}^0$ or $D^+D^-$ meson pairs. In each event, if a $D^{-}$
meson can be fully reconstructed via its decay into hadrons (in the
following called the single tag (ST) $D^{-}$), there must be a
recoiling $D^{+}$ meson. Using a double tag technique which was
first employed by the MARKIII Collaboration~\cite{mark3}, we can
measure the absolute branching fraction of the $D^{+}\rightarrow\bar
K^0e^{+}\nu_{e}$ decay. Throughout the paper, charge conjugation is
implied.

The ST $D^{-}$ mesons are reconstructed using six hadronic decay
modes: $K^{+}\pi^{-}\pi^{-}$, $K^0_{S}\pi^{-}$,
$K^{+}\pi^{-}\pi^{-}\pi^{0}$, $K^0_{S}\pi^{-}\pi^{0}$,
$K^0_{S}\pi^{+}\pi^{-}\pi^{-}$ and $K^{+}K^{-}\pi^{-}$. The daughter
particles $K^0_{S}$ and $\pi^0$ are reconstructed via $K^0_{S}\to
\pi^{+}\pi^{-}$ and $\pi^0\to\gamma\gamma$, respectively.

All charged tracks are required to be reconstructed within the good
MDC acceptance $|\cos\theta|<0.93$, where $\theta$ is the polar
angle of the track with respect to the beam direction. All tracks
except those from $K^0_S$ decays are required to originate from the
interaction region defined as $V_{xy}<1.0\;\text{cm}$ and
$|V_z|<10.0\;\text{cm}$. Here, $V_{xy}$ and $|V_{z}|$ are the
distances of closest approach to the Interaction Point (IP) of the
reconstructed track in the plane transverse to and along the beam
direction, respectively. For PID of charged particles, we combine
the $dE/dx$ and TOF information to calculate Confidence Levels for
the pion and kaon hypotheses ($CL_{\pi}$ and $CL_{K}$). A charged
track is taken as kaon (pion) if it has $CL_{K}>CL_{\pi}$
($CL_{\pi}>CL_{K}$).

The charged tracks from $K^0_{S}$ decays are required to satisfy
$|V_{z}|< 20.0\;\text{cm}$. The two oppositely charged tracks, which
are assumed as $\pi^+\pi^-$ without PID, are constrained to
originate from common vertex. A $\pi^+\pi^-$ combination is
considered as a $K^0_S$ candidate if its invariant mass lies in the
mass window $|M_{\pi^{+}\pi^{-}} - M_{K_{S}^{0}}|<
12\;\text{MeV}/c^{2}$, where $M_{K_{S}^{0}}$ is the nominal
$K^0_{S}$ mass~\cite{pdg2014}. The $\pi^{+}\pi^{-}$ combinations
with $L/\sigma_{L}> 2$ are retained, where $\sigma_{L}$ is the
uncertainty of the $K_S^0$ reconstructed decay length
$L$.

Photon candidates are selected by using the EMC
information. The shower time is required to be within
$700\;\text{ns}$ of the event start time. The shower energy is
required to be greater than 25 (50) MeV in the barrel (endcap)
region. The opening angle between the candidate shower and the
closest charged track is required to be greater than $10^{\circ}$. A
$\gamma\gamma$ combination is considered as a $\pi^0$ candidate if
its invariant mass falls in $(0.115,0.150)\;\text{GeV}/c^{2}$. To
obtain better mass resolution for the $D^-$ candidates, the
$\gamma\gamma$ invariant mass is constrained to the $\pi^{0}$
nominal mass \cite{pdg2014} via a kinematic fit.

To suppress combinatorial backgrounds, we define the variable
$\Delta E=E_{mKn\pi} - E_{\rm beam}$, which is the difference
between the measured energy of the $mKn\pi$ ($m=1,~2$, $n=1,~2,~3$)
combination ($E_{mKn\pi}$) and the beam energy ($E_{\rm beam}$). For
each ST mode, if there is more than one $mKn\pi$ combination
satisfying the above selection criteria, only the one
with the minimum $|\Delta E|$ is kept. The $\Delta E$ is required to
be within $(-25,+25)$\;MeV for the $K^{+}\pi^{-}\pi^{-}$,
$K^0_{S}\pi^{-}$, $K^0_{S}\pi^{+}\pi^{-}\pi^{-}$ and
$K^{+}K^{-}\pi^{-}$ decay modes, and be within $(-55,+40)$ MeV for
the $K^{+}\pi^{-}\pi^{-}\pi^{0}$ and $K^0_{S}\pi^{-}\pi^{0}$
combinations.

\begin{figure}[htp]
  \centering
  \includegraphics[width=2.8in]{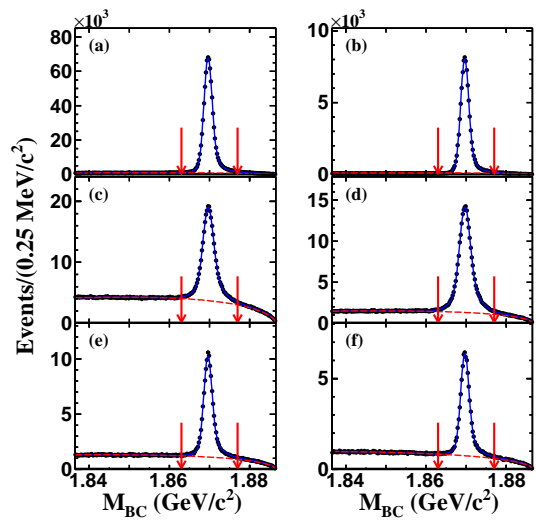}
  \caption{Fits to the $M_{\rm BC}$ spectra of the
  (a) $K^{+}\pi^{-}\pi^{-}$,
  (b) $K^0_{S}\pi^{-}$,
  (c) $K^{+}\pi^{-}\pi^{-}\pi^{0}$,
  (d) $K^0_{S}\pi^{-}\pi^{0}$,
  (e) $K^0_{S}\pi^{+}\pi^{-}\pi^{-}$ and
  (f) $K^{+}K^{-}\pi^{-}$ combinations.
The dots with error bars are data, the blue solid curves are the fit
results, the red dashed curves are the fitted
backgrounds and the pair of red arrows in each sub-figure denote
the ST $D^-$ signal region.
  }\label{fig:datafit_Massbc}
\end{figure}

To measure the yield of ST $D^{-}$ mesons, we fit the spectra of the
beam energy constrained masses ${M}_{\rm BC} = \sqrt{E^{2}_{\rm
beam}-|\vec{p}_{mKn\pi}|^{2}}$) of the accepted $mKn\pi$
combinations, as shown in Fig.~\ref{fig:datafit_Massbc}. Here,
$\vec{p}_{mKn\pi}$ is the measured momentum of the $mKn\pi$
combination. In the fits, the $D^{-}$ signal is modeled by the MC
simulated $M_{\rm BC}$ distribution convoluted with a double
Gaussian function, and the combinatorial background is described by
an ARGUS function \cite{ARGUS}. The candidates in the ST $D^-$
signal region defined as $(1.863,1.877)\;\text{GeV}/c^2$ are kept
for further analysis. Single-tag reconstruction efficiencies
$\epsilon_{\rm ST}$ are  estimated by analyzing the inclusive MC
sample. The ST yields $N_{\rm ST}$ and the ST efficiencies are
summarized in Table \ref{tab:singletagN}. The total ST yield is
$N^{\rm tot}_{\rm ST}= 1522474 \pm 2215$.

\begin{table*}[htp]
  \centering
  \caption{
   Summary of the ST yields ($N^i_{\rm ST}$),
   the ST and DT efficiencies ($\epsilon^i_{\rm ST}$ and $\epsilon^i_{\rm DT}$), and
   the reconstruction efficiencies of $D^{+}\rightarrow\bar K^0e^{+}\nu_{e}$ ($\epsilon^i_{D^{+}\rightarrow\bar K^0e^{+}\nu_{e}}$).
   The efficiencies do not include the branching fractions for
   $K^0_S\to \pi^+\pi^-$ (used in the reconstruction of ST $D^-$ mesons), $\bar K^0\to \pi^0\pi^0$ and
   $\pi^0\to\gamma\gamma$. The uncertainties are statistical only.
   The index $i$ represents the $i$th ST mode.
   }\label{tab:singletagN}
  \begin{tabular}{lcccc} \hline
  ST mode $i$ & $N^i_{\rm ST}$ & $\epsilon^i_{\rm ST}$ (\%) &
  $\epsilon^i_{\rm DT}$ (\%) &
  $\epsilon^i_{D^{+}\rightarrow\bar K^0 e^{+}\nu_{e}}$ (\%) \\
  \hline
  $D^-\to K^{+}\pi^{-}\pi^{-}$            & 782669$\pm$\hspace{0.15cm}990  & 50.61$\pm$0.06 & 13.39$\pm$0.07& 26.45$\pm$0.14\\
  $D^-\to K^0_{S}\pi^{-}$               & \hspace{0.15cm}91345$\pm$\hspace{0.15cm}320   & 50.41$\pm$0.17 & 13.81$\pm$0.22& 27.40$\pm$0.44\\
  $D^-\to K^{+}\pi^{-}\pi^{-}\pi^{0}$     & 251008$\pm$1135 & 26.74$\pm$0.09   & \hspace{0.15cm}6.23$\pm$0.06& 23.29$\pm$0.25\\
  $D^-\to K^0_{S}\pi^{-}\pi^{0}$        & 215364$\pm$1238 & 27.29$\pm$0.07  &\hspace{0.15cm}6.88$\pm$0.07& 25.21$\pm$0.28\\
  $D^-\to K^0_{S}\pi^{+}\pi^{-}\pi^{-}$ & 113054$\pm$\hspace{0.15cm}889   &28.31$\pm$0.12  &\hspace{0.15cm}6.74$\pm$0.10& 23.79$\pm$0.37\\
  $D^-\to K^{+}K^{-}\pi^{-}$              & \hspace{0.15cm}69034$\pm$\hspace{0.15cm}460   & 40.83$\pm$0.24 &10.54$\pm$0.20& 25.81$\pm$0.50\\
  \hline
\end{tabular}
\end{table*}

\subsection{Double tag events}

In the system recoiling against the ST $D^-$ mesons, the $D^+\to
\bar K^0e^+\nu_e$ candidates, called the double tag (DT) events, are
selected via $\bar K^0\to K^0_{S}\rightarrow\pi^{0}\pi^{0}$. It is
required that there be at least four good photons and only one good
charged track that have not been used in the ST selection. The good
charged track, photons and $\pi^0$ mesons are selected using the
same criteria as those used in the ST selection. If there are
multiple $\pi^0\pi^0$ combinations satisfying these selection
criteria, only the combination with the minimum value of
$\chi^2_1(\pi^0\to \gamma\gamma)+\chi^2_2(\pi^0\to \gamma\gamma)$ is
retained, where the $\chi^2_1$ and $\chi^2_2$ are the chi-squares of
the mass constrained fits on $\pi^0\to \gamma\gamma$. A
$\pi^{0}\pi^{0}$ combination is considered as a $\bar K^0$ candidate
if its invariant mass falls in $(0.45, 0.51)$ GeV/$c^{2}$. For
electron PID, we combine the $dE/dx$, TOF and EMC information to
calculate Confidence Levels for the electron, pion and kaon
hypotheses ($CL_{e}$, $CL_{\pi}$ and $CL_{K}$), respectively. The
electron candidate is required to have $CL_{e}> 0.001$ and
$CL_e/(CL_e+CL_\pi+CL_K)>0.8$, and to have a charge opposite to the
ST $D^-$ meson. To partially recover the effects of FSR and
bremsstrahlung, the four-momenta of photon(s) within $5^\circ$ of
the initial electron direction are added to the electron
four-momentum. To suppress the backgrounds associated with fake
photon(s), we require that the maximum energy ($E^{{\rm
extra}~\gamma}_{\rm max}$) of any of the extra
photons, which have not been used in the DT selection, be less than
300 MeV.

In order to obtain the information of the missing neutrino, we
define the kinematic quantity
\begin{equation}
U_{\rm miss} \equiv  E_{\rm miss} - |\vec{p}_{\rm miss}|,
\end{equation}
where $E_{\rm miss}$ and $|\vec{p}_{\rm miss}|$ are the total energy
and momentum of the missing particle in the event, respectively.
$E_{\rm miss}$ is calculated by
\begin{equation}
E_{\rm miss} = E_{\rm beam} - E_{\bar K^0} - E_{e^+},
\end{equation}
where $E_{\bar K^0}$ and $E_{e^+}$ are the energies carried by $\bar
K^0$ and $e^{+}$, respectively. $|\vec{p}_{\rm miss}|$ is calculated
by
\begin{equation}
|\vec{p}_{\rm miss}| = |\vec{p}_{D^{+}} - \vec{p}_{\bar K^0} -
\vec{p}_{e^+}|,
\end{equation}
where $\vec{p}_{D^{+}}$, $\vec{p}_{\bar K^0}$ and $\vec{p}_{e^+}$
are the momenta of $D^{+}$, $\bar K^0$ and $e^+$, respectively. To
obtain better $U_{\rm miss}$ resolution, $\vec{p}_{D^{+}}$ is
constrained by
\begin{equation}
\vec{p}_{D^{+}} = -\hat{p}_{D_{\rm ST}^{-}} \sqrt{E_{\rm
beam}^{2}-m_{D^{+}}^{2}},
\end{equation}
where $\hat{p}_{D_{\rm ST}^{-}}$ is the momentum direction of the ST
$D^{-}$ meson and $m_{D^{+}}$ is the $D^{+}$ nominal mass
\cite{pdg2014}.

\begin{figure}[htp]
  \centering
  \includegraphics[width=2.6in]{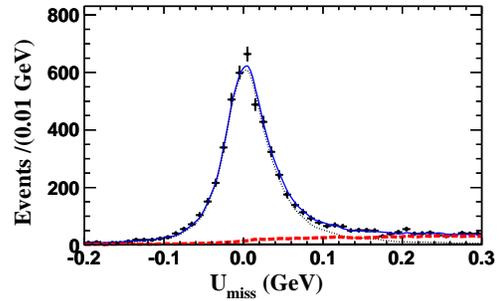}
  \caption{Fit to the $U_{\rm miss}$ distribution of the $D^{+}\rightarrow\bar K^0 e^{+}\nu_{e}$ candidates.
The dots with error bars are data, the blue solid curve is the fit
result, the black dotted and the red dashed curves are the fitted
signal and background.}\label{fig:fit_Umistry1}
\end{figure}

To determine the number of DT events, we apply a fit to the $U_{\rm
miss}$ distribution of the accepted DT candidates, as shown in
Fig.~\ref{fig:fit_Umistry1}. In the fit, the DT signal and the
combinatorial background are modeled by the MC simulated $U_{\rm
miss}$ shapes, respectively. From the fit, we obtain the DT yield in
data as
\begin{equation}
N_{\rm DT} = 5013 \pm 78.
\end{equation}

\subsection{Branching fraction}
The efficiency of reconstructing the DT events, called the DT
efficiency $\epsilon_{\rm DT}$, is determined by analyzing the
signal MC events. Dividing $\epsilon_{\rm DT}$ by $\epsilon_{\rm
ST}$, we obtain the reconstruction efficiency for
$D^{+}\rightarrow\bar K^0 e^{+}\nu_{e}$ in each ST mode,
$\epsilon_{D^{+}\rightarrow\bar K^0 e^{+}\nu_{e}}$, as summarized in
Table \ref{tab:singletagN}. Weighting them by the ST yields observed
in data, we obtain the averaged reconstruction efficiency of
$D^{+}\rightarrow\bar K^0e^{+}\nu_{e}$
\begin{equation}
\bar  \epsilon_{D^{+}\rightarrow\bar K^0 e^{+}\nu_{e}} =
(25.58\pm0.11)\%,
\end{equation}
which does not include the branching fractions of $\bar K^0\to
\pi^0\pi^0$ and $\pi^0\to\gamma\gamma$.

The branching fraction of $D^{+}\rightarrow\bar K^0e^{+}\nu_{e}$ is
determined by
 \begin{eqnarray}
   &&{\mathcal B}(D^{+}\rightarrow \bar K^0e^{+}\nu_{e}) \nonumber
   = \\
   &&\frac{N_{\rm DT}}
   {N^{\rm tot}_{\rm ST} \bar \epsilon_{D^{+}\rightarrow\bar K^0e^{+}\nu_{e}}
   {\mathcal B}(\bar K^0\rightarrow\pi^0\pi^0)
   {\mathcal B}^2(\pi^0\rightarrow\gamma\gamma)
   },
\label{equ:bf}
 \end{eqnarray}
where $N_{\rm DT}$ is the DT yield, $N^{\rm tot}_{\rm ST}$ is the
total ST yield, $\bar \epsilon_{D^{+}\rightarrow\bar
K^0e^{+}\nu_{e}}$ is the averaged reconstruction efficiency of
$D^{+}\rightarrow\bar K^0e^{+}\nu_{e}$, ${\mathcal B}(\bar
K^0\rightarrow\pi^0\pi^0)$ and ${\mathcal
B}(\pi^0\rightarrow\gamma\gamma)$ are the branching fractions of
$\bar K^0\rightarrow\pi^{0}\pi^{0}$ and
$\pi^{0}\rightarrow\gamma\gamma$~\cite{pdg2014}, respectively. Here,
we assume that $K^0_S$ constitutes half the decays of the neutral
kaons.

Inserting the numbers of $N_{\rm DT}$, $N^{\rm tot}_{\rm ST}$, $\bar
\epsilon_{D^{+}\rightarrow\bar K^0e^{+}\nu_{e}}$, ${\mathcal B}(\bar
K^0\rightarrow\pi^0\pi^0)$
 and ${\mathcal B}(\pi^0\rightarrow\gamma\gamma)$
in Eq.~(\ref{equ:bf}), we obtain
$${\mathcal B}(D^{+}\rightarrow\bar K^0e^{+}\nu_{e}) = (8.59 \pm 0.14)\%,$$
where the uncertainty is statistical only.

\subsection{Systematic uncertainty}
In the measurement of the branching fraction, the systematic
uncertainty arises from the uncertainties in the fits to the $M_{\rm
BC}$ spectra of the ST candidates, the $\Delta E$, $M_{\rm BC}$ and
$\bar K^0(\pi^0\pi^0)$ mass requirements, the $\pi^0$
reconstruction, the $e^\pm$ tracking, the $e^\pm$ PID, the $E^{\rm
extra~\gamma}_{\rm max}$ requirement, the $U_{\rm miss}$ fit, the
$\chi^2_1+\chi^2_2$ selection method, the MC statistics and the
quoted branching fractions.

The uncertainty in the fits to the $M_{\rm BC}$ spectra of the ST
candidates is estimated to be 0.5\% by observing the relative change
of the ST yields of data and MC when varying the fit range, the
combinatorial background shape or the endpoint of the ARGUS
function. To estimate the uncertainties in the $\Delta E$, $M_{\rm
BC}$ and $\bar K^0(\pi^0\pi^0)$ mass requirements, we examine the
change in branching fractions when enlarging the $\Delta E$
selection window by 5 or 10 MeV; varying the $M_{\rm BC}$ selection
window by $\pm1\;\text{MeV}$ and using alternative $\bar
K^0(\pi^0\pi^0)$ mass windows $(0.460, 0.505)$, $(0.470, 0.500)$,
$(0.480, 0.500)\;\text{GeV}/c^2$, respectively. The maximum changes
in the branching fractions, 0.3\%, 0.2\%, and 0.9\%, are assigned as
the systematic uncertainties. The $\pi^0$ reconstruction efficiency
is examined by analyzing the DT hadronic decays $D^0\to K^-\pi^+$
and $K^-\pi^+\pi^+\pi^-$ versus $\bar{D^0}\to K^-\pi^+\pi^0$ and
$K_{S}^{0}(\pi^+\pi^-)\pi^0$. The difference of the $\pi^0$
reconstruction efficiencies between data and MC is found to be
$(-1.0\pm1.0)\%$ per $\pi^0$. The systematic uncertainty in $\pi^0$
reconstruction is taken to be $1.0\%$ for each $\pi^0$ after
correcting the MC efficiency of $D^+\to \bar K^0e^+\nu_e$ to data.
The uncertainty in the tracking or PID for $e^\pm$ is estimated by
analyzing $e^+e^-\to \gamma e^+e^-$ events. It is assigned to be
0.5\%, which is the re-weighted difference of the $e^\pm$ tracking
(or PID) efficiencies between data and MC. The uncertainty in the
$E_{\rm max}^{\rm extra~\gamma}$ requirement is estimated to be
0.1\% by analyzing the DT hadronic $D\bar{D}$ decays. The
uncertainty in the $U_{\rm miss}$ fit is assigned to be 0.5\%, which
is obtained by comparing with the nominal value of the
branching fraction measured with an alternative signal shape
obtained with different requirements on the MC-truth matched signal
shape, an alternative background shape after changing the relative
ratios of the dominant backgrounds (doubling each of the simulated
backgrounds for $D^0\bar D^0$, $D^+D^-$ and $q\bar q$ continuum
processes), and alternative fit range ($\pm 50\;\text{MeV}$). The
difference of 0.3\% in the $\pi^0\pi^0$ acceptance efficiencies
between data and MC, which is estimated by the DT hadronic decays
$D^0\to K^-\pi^+\pi^0$ versus $\bar D^0\to K^+\pi^-\pi^0$, is
assigned as a systematic uncertainty due to the $\chi^2_1+\chi^2_2$
selection method. In this analysis, the $\bar K^0\to
K^0_S(\pi^0\pi^0)$ meson from the signal side is formed with photon
candidates reconstructed under the assumption that they originate at
the IP. We examine the DT efficiencies of the signal MC events in
which the lifetimes of $K^0_S$ meson from the signal side are set to
the nominal value and 0, respectively. The difference of these two
DT efficiencies, which is less than 0.2\%, is taken as the
systematic uncertainty of the $K^0_S(\pi^0\pi^0)$ reconstruction.
The uncertainties in the MC statistics and the ${\mathcal B}(\bar
K^0 \to \pi^0\pi^0)$ are 0.5\% and 0.2\%~\cite{pdg2014},
respectively. In our previous work, the uncertainty in the signal MC
generator is estimated to be 0.1\%, which is obtained by comparing
the DT efficiencies before and after re-weighting the
$q^2(=(p_D-p_K)^2)$ distribution of the signal MC events of $D^0\to
K^-e^+\nu_e$ to the distribution found in data~\cite{bes3-kev},
where the $p_D$ and $p_K$ are the four-momenta of the $D$ and $K$
mesons. The systematic uncertainties are summarized in
Table~\ref{tab:sys}. Adding all uncertainties in quadrature, we
obtain the total systematic uncertainty to be 2.5\%.

\begin{table}[htp]
\centering \caption{Relative systematic uncertainties (in \%) in the
measurement of ${\mathcal B}(D^{+}\rightarrow\bar K^{0}
e^{+}\nu_e)$.} \label{tab:sys}
\begin{tabular}{cc}
  \hline
  Source & Uncertainty \\
  \hline
  $M_{\rm BC}$ fit &  0.5 \\
  $\Delta E$ requirement &  0.3 \\
  $M_{\rm BC}\in(1.863,1.877)$ GeV/$c^2$ &  0.2 \\
  $M_{\pi^0\pi^0}\in(0.45, 0.51)$ GeV/$c^{2}$ & 0.9\\
  $\pi^0$ reconstruction &  2.0 \\
  Tracking for $e^\pm$ &  0.5 \\
  PID for $e^\pm$ &  0.5 \\
  $E^{\rm extra~\gamma}_{\rm max}<0.3$ GeV &  0.1 \\
  $U_{\rm miss}$ fit &  0.5 \\
  $\chi^2_1+\chi^2_2$ selection method& 0.3 \\
  $K^0_S(\pi^0\pi^0)$ reconstruction & 0.2 \\
  MC statistics &  0.5 \\
  ${\mathcal B}(\bar K^0 \to \pi^0\pi^0)$ &  0.2 \\
  MC generator  &  0.1 \\ \hline
  Total & 2.5 \\ \hline
\end{tabular}
\end{table}

\subsection{Validation}
The analysis procedure is examined by an input and output check
using an inclusive MC sample equivalent to a luminosity of 3.26
fb$^{-1}$. Using the same selection criteria as those used in data
analysis, we obtain the ST yield, the DT yield and the weighted
reconstruction efficiency of $D^{+}\rightarrow\bar K^0 e^{+}\nu_{e}$
to be $1683631\pm1768$, $5802\pm85$ and $(26.07\pm0.11)\%$, where no
efficiency correction has been performed. Based on these numbers, we
determine the branching fraction ${\mathcal B}(D^{+}\rightarrow\bar
K^0 e^{+}\nu_{e})=(8.82\pm0.13)\%$, where the uncertainty is
statistical only. The measured branching fraction is in excellent
agreement with the input value of 8.83\%.

\begin{figure}[htp]
  \centering
  \includegraphics[width=2.8in]{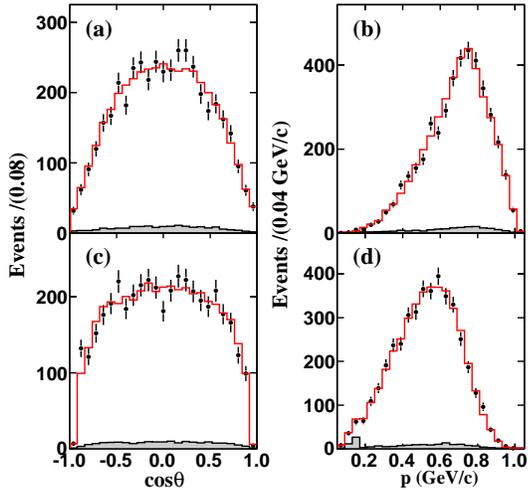}
  \caption{Comparisons of the $\cos\theta$ and momentum distributions of
  (a), (b) $\bar K^0$ and (c), (d) $e^+$ of
  the $D^{+}\rightarrow\bar K^0 e^{+}\nu_{e}$ candidates.
  The dots with error bars are data,
  the red histograms are the inclusive MC events, and the
  light black hatched histograms are the MC
  simulated backgrounds. These events satisfy a tight requirement of $-0.06<U_{\rm miss}<+0.06\;\text{GeV}$.
}\label{fig:comparison}
\end{figure}

To validate the reliability of the MC simulation, we examine the
$\cos\theta$ and momentum distributions of $\bar K^0$ and $e^+$ of
the $D^{+}\rightarrow\bar K^0 e^{+}\nu_{e}$ candidates, as shown in
Fig.~\ref{fig:comparison}. We can see that the consistency between
simulation and data is very good.

\section{Summary and discussion}

\begin{figure}[htp]
  \centering
  \includegraphics[width=2.8in]{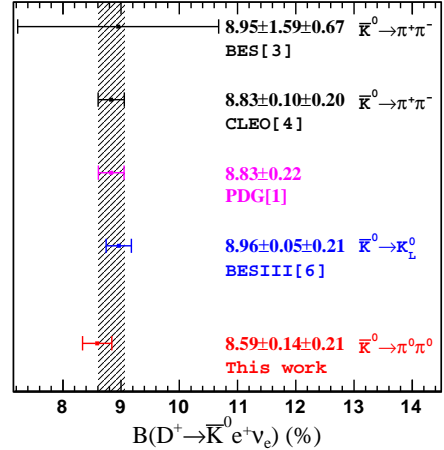}
  \caption{
  Comparison of the ${\mathcal B}(D^{+}\rightarrow\bar K^0 e^{+}\nu_{e})$
  measured in this work with those measured by other experiments, where
  the slash band is the world averaged branching fraction with uncertainty.
  For the BESIII measurement using $\bar K^0\to K^0_L$, we take
  ${\mathcal B}(D^{+}\rightarrow\bar K^0 e^{+}\nu_{e})=2{\mathcal B}(D^{+}\rightarrow K^0_L e^{+}\nu_{e})$.
  }
  \label{fig:bf-compare}
\end{figure}

Based on the analysis of 2.93 fb$^{-1}$ data collected at $\sqrt{s}=
3.773\;\text{GeV}$ with the BESIII detector, we measure the absolute
branching fraction ${\mathcal B}(D^{+}\rightarrow\bar
K^0e^{+}\nu_{e}) = (8.59 \pm 0.14 \pm 0.21)\%$, using $\bar K^0\to
K^0_S\to\pi^0\pi^0$. Figure~\ref{fig:bf-compare} presents a
comparison of ${\mathcal B}(D^{+}\rightarrow\bar K^0 e^{+}\nu_{e})$
measured in this work with the results obtained by other
experiments. Our result is well consistent with the other
measurements within uncertainties and has a precision
comparable to the PDG value~\cite{pdg2014}. Our measurement will be
helpful to improve the precision of the world average value of
${\mathcal B}(D^+\to \bar K^0 e^{+}\nu_{e})$.

Combining the PDG values for ${\mathcal B}(D^0\to K^-e^+\nu_e)$,
${\mathcal B}(D^{+}\rightarrow\bar K^0e^{+}\nu_{e})$ \cite{pdg2014},
and the lifetimes of $D^0$ and $D^+$ mesons ($\tau_{D^0}$ and
$\tau_{D^+}$)~\cite{pdg2014} with the value of ${\mathcal
B}(D^{+}\rightarrow\bar K^0e^{+}\nu_{e})$ measured in this work, we
determine
\begin{eqnarray}
\frac{\Gamma(D^0\to K^-e^+\nu_e)}{\bar \Gamma(D^{+}\rightarrow\bar
K^0e^{+}\nu_{e})} \nonumber &&=
\frac{{\mathcal B}(D^0\to K^-e^+\nu_e)\times \tau_{D^+}}{\bar {\mathcal B}(D^{+}\rightarrow\bar K^0e^{+}\nu_{e})\times\tau_{D^0}}\nonumber \\
&&=0.969\pm0.025,
\end{eqnarray}
where $\bar {\mathcal B}(D^{+}\rightarrow\bar K^0e^{+}\nu_{e})$ is
the the averaged branching fraction based on the PDG value and the
one measured in this work. This gives a more stringent test on
isospin symmetry in the $D^+\to \bar K^0e^+\nu_e$ and $D^0\to
K^-e^+\nu_e$ decays.

\section{Acknowledgements}
The BESIII collaboration thanks the staff of BEPCII and the IHEP
computing center for their strong support. This work is supported in
part by National Key Basic Research Program of China under Contract
Nos. 2009CB825204 and 2015CB856700; National Natural Science
Foundation of China (NSFC) under Contracts Nos. 10935007, 11125525,
11235011, 11305180, 11322544, 11335008, 11425524, 11475123; the
Chinese Academy of Sciences (CAS) Large-Scale Scientific Facility
Program; the CAS Center for Excellence in Particle Physics (CCEPP);
the Collaborative Innovation Center for Particles and Interactions
(CICPI); Joint Large-Scale Scientific Facility Funds of the NSFC and
CAS under Contracts Nos. 11179007, U1232201, U1332201, U1532101; CAS
under Contracts Nos. KJCX2-YW-N29, KJCX2-YW-N45; 100 Talents Program
of CAS; National 1000 Talents Program of China; INPAC and Shanghai
Key Laboratory for Particle Physics and Cosmology; German Research
Foundation DFG under Contract No. Collaborative Research Center
CRC-1044; Istituto Nazionale di Fisica Nucleare, Italy; Koninklijke
Nederlandse Akademie van Wetenschappen (KNAW) under Contract No.
530-4CDP03; Ministry of Development of Turkey under Contract No.
DPT2006K-120470; National Natural Science Foundation of China (NSFC)
under Contracts Nos. 11405046, U1332103; Russian Foundation for
Basic Research under Contract No. 14-07-91152; The Swedish Resarch
Council; U. S. Department of Energy under Contracts Nos.
DE-FG02-04ER41291, DE-FG02-05ER41374, DE-SC0012069, DESC0010118;
U.S. National Science Foundation; University of Groningen (RuG) and
the Helmholtzzentrum fuer Schwerionenforschung GmbH (GSI),
Darmstadt; WCU Program of National Research Foundation of Korea
under Contract No. R32-2008-000-10155-0.

\end{document}

%% file: BESIII_authors_en.tex
\author{
M.~Ablikim$^{1}$, M.~N.~Achasov$^{9,e}$, X.~C.~Ai$^{1}$,
O.~Albayrak$^{5}$, M.~Albrecht$^{4}$, D.~J.~Ambrose$^{44}$,
A.~Amoroso$^{49A,49C}$, F.~F.~An$^{1}$, Q.~An$^{46,a}$,
J.~Z.~Bai$^{1}$
% ...
\\
\vspace{0.2cm}
(BESIII Collaboration)\\
\vspace{0.2cm} {\it
$^{1}$ Institute of High Energy Physics, Beijing 100049, People's Republic of China\\
% ...
$^{4}$ Bochum Ruhr-University, D-44780 Bochum, Germany\\
$^{5}$ Carnegie Mellon University, Pittsburgh, Pennsylvania 15213, USA\\
% ...
$^{9}$ G.I. Budker Institute of Nuclear Physics SB RAS (BINP), Novosibirsk 630090, Russia\\
% ...
$^{44}$ University of Rochester, Rochester, New York 14627, USA\\
% ...
$^{46}$ University of Science and Technology of China, Hefei 230026, People's Republic of China\\
% ...
$^{49}$ (A)University of Turin, I-10125, Turin, Italy; (B)University of Eastern Piedmont, I-15121, Alessandria, Italy; (C)INFN, I-10125, Turin, Italy\\
% ...
$^{a}$ Also at State Key Laboratory of Particle Detection and Electronics, Beijing 100049, Hefei 230026, People's Republic of China\\
% ...
$^{e}$ Also at the Novosibirsk State University, Novosibirsk, 630090, Russia\\
% ...
}
}